\documentclass[12pt]{article}
\usepackage{graphicx}
\usepackage{cite}
\usepackage{amssymb,amsmath}
\begin{document}
\title{Blocking of DNA specific recognition sites by hydrogen peroxide molecules in the process of ion beam therapy of cancer cells}%
\author{Oleksii Zdorevskyi, Dmytro Piatnytskyi, Sergey N. Volkov\\
Bogolyubov Institute for Theoretical Physics, NAS of
Ukraine,\\14-b Metrolohichna Str., Kiev 03143, Ukraine \\
snvolkov@bitp.kiev.ua }\maketitle
\setcounter{page}{1}%
\maketitle
\begin{abstract}
	After irradiation of cancer cells in the ion beam therapy method the concentration of hydrogen peroxide in the cell medium grows significantly. But the role of hydrogen peroxide molecules in cancer treatment has not been determined yet. We assume that interaction of peroxide molecules with DNA atomic groups can block the genetic information of the cancer cell and lead to its neutralization. To understand the possibility of DNA deactivation in the cell, in the present study the formation of complexes of hydrogen peroxide molecules with DNA specific recognition sites (nucleic bases) is considered. Using atom-atom potential functions method and quantum-chemical approach, based on density functional theory, the spatial configurations and energy minima for the complexes of peroxide and water molecules with nucleic bases are studied. The most probable positions of hydrogen peroxide molecules interacting with nucleic bases are determined, and the possibility of blocking of genetic information transfer processes is shown. The obtained data allows us to formulate a new mechanism of the ion irradiation action on living cells, that can be useful for cancer treatment.

\end{abstract}

\section{Introduction}

DNA macromolecule under physiological conditions forms a double helix consisting of two chains
of a sugar-phosphate backbone, which are interconnected by hydrogen bonds. The medium of the biological cell mainly
consists of water molecules and positively charged alkali metal ions that neutralize negatively charged phosphate
groups of DNA backbone. The water content, the concentration of counterions and the cell solution, determine the form of the double helix, and consequently influence significantly on DNA functioning.\cite{Saenger} 

At physiological conditions all the main biological processes in living cell are determined by the information which is stored in DNA macromolecule. But in the cancer cells the processes of genetic information transfer take place without any appropriate control. This is considered as the main reason of the tumor appearance. Therefore, the relevant task of cancer treatment is studying of the mechanisms of  deactivation of cancer cells by the direct destruction of their DNA in a cell or by the blocking of active DNA recognition sites.

One of the most progressive methods for cancer treatment is ion beam therapy, when  the
living tissue is irradiated with high-energy ion beams. The beam is formed on accelerator, and it is used for the treatment of patients at the special hospitals.\cite{KramerDurante2010} In the basics of ion beam therapy lies the well-known effect of Bragg's
peak formation\cite{Kraft2000}. It is considered, that the main target for high-energy ion beam in the cell is DNA, but
a current mechanism of the action of heavy ions on DNA of cancer cells has not been determined yet.\cite{KramerDurante2010}

As the DNA macromolecule in the cell nucleus interacts with the water
medium all the time, the influence of ion beam on water solution of the cell actually determines the state of DNA macromolecule and the flow of genetic processes as well.
The study of water radiolysis\cite{Kreipl2008,Uehara2006} showed that under the influence of ionizing radiation in the water medium different species occur, such as secondary electrons, ions, free radicals, as well as molecular products ($H_2$, $H_2O_2$). According to the traditional hypothesis of ion therapy, it is considered
that the main reason of DNA deactivation are the double-strand breaks caused by secondary electrons and free radicals.\cite{Kraft2000}
However, it is now known\cite{NobelPrize2015} that there are powerful mechanisms of DNA repair, which restore the functioning of the
molecule after double-strand breaks. Therefore, it is necessary to study also other mechanisms of DNA deactivation, particularly the processes of blocking of the active DNA recognition cites by the molecular products of radiolysis. 

Monte Carlo simulations and experiments on water radiolysis\cite{Kreipl2008,Uehara2006,Durante2018} 
showed that at the scale of biological lifetimes ($\sim 1\mu $sec), among all species hydrogen peroxide
molecules ($H_2O_2$) have the largest concentration. Taking this into account, one of the mechanisms of the ionizing radiation action on DNA, that is proposed in the paper,\cite{pyatEPJ2015} may be the  blocking of the active protein-DNA recognition sites by the hydrogen peroxide molecules. In normal conditions, these sites are surrounded by water molecules. But due to increase of $H_2O_2$ concentration, these molecules should compete with water molecules for binding with DNA sites. Therefore, it is important to understand whether hydrogen peroxide molecules can interact with active DNA recognition sites more stronger than water molecules, thus blocking the processes of genetic information transfer.

As known, at physiological conditions, hydrogen peroxide molecules are present in living cell at the certain concentration, but their role in the cell's functioning has not been studied properly yet.\cite{watsonMolBiolGene} There are some methods of cancer therapy (see for example\cite{levine2013}), where hydrogen peroxide molecules are introduced into the cellular medium and selectively damage the cancer cells.\cite{levine2013,chen2005} But the mechanisms of this treatment have not been determined yet.

In our work\cite{pyatEPJ2015} the interaction of hydrogen peroxide molecules with centers of
non-specific protein-nucleic recognition - DNA phosphate groups ($PO_4$) was studied. It was shown that $H_2O_2$ molecule can form a stable, long-living complex with DNA $PO_4$ groups, and block the protein-nucleic recognition process. In the present paper, we consider the interaction of $H_2O_2$ and $H_2O$ molecules with specific DNA recognition sites - atomic groups of Adenine, Thymine, Guanine, and Cytosine nucleic bases, to determine if they can form stable complexes. Since water and hydrogen peroxide molecules have a similar atomic content, they must compete in the solution for the binding with DNA.

The next Section will describe our research methods. In Sec. \ref{calcRes}, a comparative analysis of the stability of the complexes  consisting of a $H_2O_2$ and $H_2O$ molecules with nucleic base (Adenine, Thymine, Guanine
or Cytosine) will be performed. In Sec. \ref{discuss} it will be discussed how our results can lead to the understanding of the mechanism of blocking of the genetic information transfer processes.

\section{Calculation methods}
\label{calcMet}
For the analysis of interaction energy and structure of the investigated molecular complexes two computational approaches are used - the method of classical atom-atom potential functions (AAPF) and the method of quantum-chemical calculations based on density functional theory with B3LYP functional.

\begin{figure}
	\begin{center}
		\resizebox{1\textwidth}{!}{%
			\includegraphics{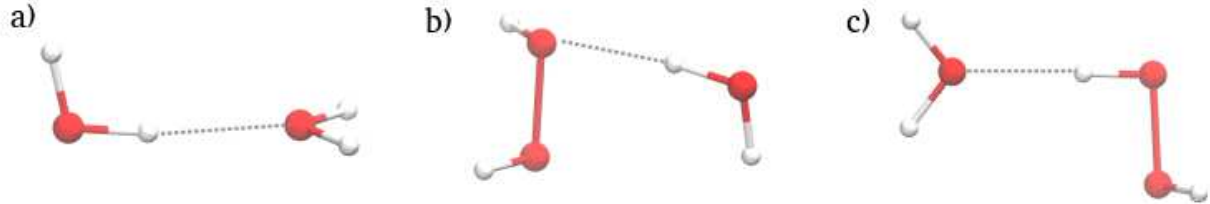}
		}
		\caption{Spatial configurations for complexes consisting of: a) two water molecules; b), c)
			hydrogen peroxide molecule and a water molecule.}
		\label{fig:H2OH2O2}       % Give a unique label
	\end{center}
\end{figure}

In this Section we will describe these two approaches and show how they agree with the previously calculated data for the simple complexes consisting of water and hydrogen peroxide molecules. In this way the energy minima and spatial configurations for molecular complexes are found (Fig. \ref{fig:H2OH2O2}).

\subsection{Atom-atom potential functions method}

The atom-atom potential function method is now widely used in the molecular dynamics in such force fields as CHARMM and AMBER\cite{Charmm,Amber,LaveryMolDynRev} for studying the structure of molecular complexes. In the framework of this method, the energy of intermolecular interaction consists of van der Waals interactions, hydrogen bonds and and Coulomb interactions:
\begin{equation}
E\left(r\right)=\sum
_{i,j}\left(E_{\mathit{vdW}}\left(r_{\mathit{ij}}\right)+E_{\mathit{HB}}\left(r_{\mathit{ij}}\right)+E_{\mathit{Coul}}\left(r_{\mathit{ij}}\right)\right).
\end{equation}

In the framework of the present method we consider all the covalent bonds and angles as rigid.

Van der Waals's interaction is described by Lennard-Jones's '6-12' potential:
\begin{equation}
E_{\mathit{vdW}}\left(r_{\mathit{ij}}\right)=-\frac{A_{\mathit{ij}}}{r_{\mathit{ij}}^6}+\frac{B_{\mathit{ij}}}{r_{\mathit{ij}}^{12}},
\end{equation}
where the parameters
$A_{ij}^{(10)}$, $B_{ij}^{(10)}$, $A_{ij}$, $B_{ij}$ are taken from the works.\cite{PoltevShul1986,Poltev1980}

The energy of the hydrogen bond between atoms $i$ and $j$ is modeled by the modified Lenard-Jones potential '10 -12':

\begin{equation}
E_{\mathit{HB}}\left(r_{\mathit{ij}}\right)=\left[-\frac{A_{\mathit{ij}}^{\left(10\right)}}{r_{\mathit{ij}}^{10}}+\frac{B_{\mathit{ij}}^{\left(10\right)}}{r_{\mathit{ij}}^{12}}\right]\cos
\varphi ,
\end{equation}
where $r_{ij}$ - the distance
between the atoms $i$ and $j$,  $\varphi $ - the angle of the hydrogen bond. For example, when the hydrogen bond is $O-H
... N$, then  $\varphi $  is an angle between the lines of covalent bond ($O-H$) and the hydrogen bond ($H ... N$).

Coulomb interaction is described by the electrostatic potential:
\begin{equation}
E_{\mathit{Coul}}\left(r_{\mathit{ij}}\right)=\frac 1{4\pi \varepsilon _0\varepsilon
	\left(r_{\mathit{ij}}\right)}\frac{q_iq_j}{r_{\mathit{ij}}},
\end{equation}
where $q_i$ and $q_j$ are the charges of the atoms $i$ and $j$ located at a distance $r_{ij}$, $\varepsilon_0$ is the vacuum permittivity, and $\varepsilon(r)$ is the dielectric permittivity of the medium.

The charges $q_i$, $q_j$ for nucleic bases were taken from the works.\cite{PoltevShul1986,Poltev1980} Charges of $H_2O$ and $H_2O_2$ molecules were calculated from the condition that the dipole moment of water molecule should be equal to $d_{H2O}=1.86 D$, and of hydrogen peroxide molecule $d_{H2O2} = 2.10 D$.\cite{pyatEPJ2015} Hence, for the $H_2O$ molecule we obtain the charges $q_H = 0.33e$, $q_O = -0.66e$, and, accordingly, for $H_2O_2$ $q_H = 0.41e, q_O = -0.41e$. The values of charges on the atoms of $H_2O_2$ molecule are in good agreement with charges obtained in the work.\cite{Moin2012}

Since DNA in the living cell is situated in a water-ion solution, the interacting atoms are
screened by water molecules. This leads to a weakening of the Coulomb interaction. Thus, more effective accounting of Coulomb interactions can be achieved using the dependence of the dielectric permittivity upon distance ($\varepsilon $(r)), developed by Hingerty \textit{et al.}\cite{hingerty} in the explicit form:

\begin{equation}
\label{for:hing}
\epsilon \left(r\right)=78-77\left(r_p\right)^2\frac{e^{r_p}}{\left(e^{r_p}-1\right)^2},
\end{equation}
where  $r_p=r/2.5.$  Further atom-atom potential functions method (AAPF) with the use of expression (\ref{for:hing}) will be called here as AAPFh.

\subsection{Quantum chemistry approach}
In the quantum-chemical calculations the B3LYP/6-311+G(d,p) method with a supermolecular approach is used. Calculations are made within Gaussian software.\cite{gaussian} The interaction energy is considered as a difference between energy of the molecular complex $XY$ and its components $X$ and $Y$:

\begin{equation}
\Delta E_{XY}= E_{XY}(XY)-E_X(X)-E_Y(Y).
\end{equation}
The basis sets of the molecular complex XY (dimer centered basis set) and of the isolated molecules X and Y (monomer centered basis sets) are shown in parenthesis. The counterpoise correction (CP) is made to avoid basis set superposition error, as following:
\begin{equation}
\Delta E^{CP}_{XY}=E_{XY}(XY)-E_X(XY)-E_Y(XY).
\end{equation}

In the framework of this approach we calculate deformation energy for hydrogen peroxide and water molecules, because we assume that nucleic bases are the rigid structures.  Deformation energy is calculated to take into account the possible differences between the isolated states of the molecules and of those within complexes:

\begin{equation}
E_{def}=E_{complex}-E_{isolated}.
\end{equation}
Thus, complete interaction energy in complex is presented by formula:

\begin{equation}
\Delta E_{complete}=\Delta E^{CP}+E_{def}.
\end{equation}

In addition in the framework of the present method, to get an electrostatically neutral structure, atoms $C_1'$ that form glycosidic bonds of the nucleic base with DNA backbone are changed to $H$ atoms for  Adenine and Thymine and to $CH_3$ group for Guanine and Cytosine.

\subsection{Test calculations of simple complexes with hydrogen bonding}
\label{comparison}
Using AAPF and B3LYP methods the interaction energies of the considered complexes consisting of $H_2O_2$ and $H_2O$ molecules (Fig. \ref{fig:H2OH2O2}) are calculated and compared with the corresponding data from the works.\cite{Kumar2014,Moin2012,Du2005,gonzalez1997} 

As can be seen from the results of our calculations (Tabls. \ref{tab:H2OH2O} and \ref{tab:H2OH2O2}), in the framework of AAPFh approach the energy minima values, as well as spatial configurations of the molecules, are consistent with the corresponding works much better than the same calculations carried out in a gas phase. Also in the work\cite{ourWorkUnz} it was shown the necessity of use of dependence (\ref{for:hing}) when calculating interactions in nucleic base pairs to avoid anomalous Coulomb contributions. Due to these facts, all the following calculations in the framework of AAPF method we perform taking into account  the dependence (\ref{for:hing}).

\begin{table}[th]
\begin{center}	
	\noindent\caption{Comparison of spatial configurations and interaction energies in water-water (Fig. \ref{fig:H2OH2O2}a) complex, calculated in this paper using atomic-atom potential functions (AAPF) method and the method of quantum chemistry B3LYP, as well as data calculated for the same complexes in other works. Distances are given in ${\AA}$, angles in degrees, energies in $kcal/mol$.
		\label{tab:H2OH2O}}
	{\begin{tabular}{@{}ccccccc@{}} %\toprule
			& & & & & & \\
			\multicolumn{7}{c}{\textbf{$H_2O-H_2O$ $(a)$}} \\  \hline
			& & \multicolumn{3}{c}{Our calculations} & & taken from\cite{Kumar2014} \\  
			Method & & AAPF & AAPFh & B3LYP & & B3LYP \\ \hline
			& & & & & & \\
			$O...O$ & & $2.79$ & $2.79$ & $2.90$ & & $2.94$\\
			$H...O$ & & $1.84$ & $1.84$ & $1.93$ & & $1.98$\\
			$Angle$ & & $173.2$ & $173.8$ & $173.8$ & & $173.1$\\
			$E$ & & $-7.25$ & $-5.71$ & $-5.08$ & & $-5.08$\\  \hline
	\end{tabular}}
\end{center}
\end{table}

\begin{table}[th]
	\begin{center}	
		\noindent\caption{Comparison of spatial configurations and interaction energies in peroxide-water (Fig. \ref{fig:H2OH2O2}b) and (Fig. \ref{fig:H2OH2O2}c) complexes, calculated in this paper using atomic-atom potential functions (AAPF) method and the method of quantum chemistry B3LYP, as well as data calculated for the same complexes in other works. Distances are given in ${\AA}$, angles in degrees, energies in $kcal/mol$.
			\label{tab:H2OH2O2}}
		
		{\begin{tabular}{@{}ccccccccc@{}} 
				\multicolumn{9}{c}{\textbf{$H_2O_2-H_2O$  $(b)$}} \\  \hline
				& & \multicolumn{3}{c}{Our calculations} & & taken from\cite{Moin2012} & taken from\cite{Du2005} & taken from\cite{gonzalez1997} \\  
				Method & & AAPF & AAPFh & B3LYP & & B3LYP & B3LYP & B3LYP \\	\hline
				&&&&&&& \\
				$O...O$ & & $2.73$ & $2.80$ & $2.97$ & & ---  & --- & $2.97$\\
				$H...O$ & & $1.86$ & $1.85$ & $2.01$ & & $1.99$ & $2.03$ & $2.00$\\
				$Angle$ & & $149.2$ & $169.7$ & $174.2$ & & $168.5$ & $170.1$ & ---\\
				$E$ & & $-6.95$ & $-5.77$ & $-3.90$ & & $-5.10$ & $-3.56$ & $-2.70$\\  \hline
				&&&&&&&& \\
				\multicolumn{9}{c}{\textbf{$H_2O_2-H_2O$  $(c)$}} \\  \hline
				& & \multicolumn{3}{c}{Our calculations} & & taken from\cite{Moin2012} & taken from\cite{Du2005} & taken from\cite{gonzalez1997} \\  
				Method & & AAPF & AAPFh & B3LYP & & B3LYP & B3LYP & B3LYP \\	\hline
				&&&&&&& \\
				$O...O$ & & $2.78$ & $2.78$ & $2.78$ & & ---  & --- & $2.78$\\
				$H...O$ & & $1.82$ & $1.82$ & $1.89$ & & $1.86$ & $1.94$ & $1.91$\\
				$Angle$ & & $178.1$ & $178.1$ & $148.5$ & & $145.9$ & $146.2$ & ---\\
				$E$ & & $-9.17$ & $-7.00$ & $-6.37$ & & $-9.60$ & $-6.31$ & $-5.10$\\  \hline
		\end{tabular}}
		\end{center}
        \end{table}

In quantum-chemical approach (B3LYP), the obtained energy values for the water-water complex coincide quite good with the previously calculated data. As it can be seen from Tabl. \ref{tab:H2OH2O}, the obtained values of the interaction energy in gas phase, as well as the geometry of the complex, are very close to the data obtained in the paper.\cite{Kumar2014} Also our calculated data is in a good accordance with values from the review.\cite{Mukhopadhyay2018} 

For the system peroxide-water (Tabl. \ref{tab:H2OH2O2}), the calculated spatial configurations are also in good agreement with the works.\cite{Moin2012,Du2005,gonzalez1997} The calculated energies of the complexes are very close to the results obtained in the work.\cite{Du2005}

To sum up, the used approaches (AAPFh and B3LYP) give sufficiently realistic results and can be used for the analysis of the interactions of $H_2O_2$ and $H_2O$ molecules with nucleic bases.

\section{Studying of the competitive interactions of $H_2O$ and $H_2O_2$ molecules with atomic groups of nucleic bases}
\label{calcRes}
The interaction of nucleic bases with water molecules has been considered in the set of works,\cite{Kryachko,Lavery2003,Clementi1980,PoltevWater} but the interaction with hydrogen peroxide molecules has not been studied sufficiently yet. Up to now only the work\cite{dobado1999} is known.
In the present work, the stable complexes of
hydrogen peroxide molecules with nucleic bases A, T, G and C are found and compared to the same complexes with the water molecule.

It is clear, that hydrogen peroxide, as well as water, can form hydrogen bonds with nucleic bases. Due to the structures of hydrogen peroxide molecule and nucleic base, it is much more favorable for peroxide to form two hydrogen bonds with base atomic groups, where
two different oxygen atoms of the $H_2O_2$ molecule are involved in the formation of hydrogen bonds. Moreover,
unlike a complex with one hydrogen bond, in a complex with two hydrogen bonds, the peroxide molecule should be more stable, it
has no possibility to rotate around hydrogen bonds. Therefore, in accordance with the geometry of water and hydrogen peroxide molecules, in this paper we
consider only the complexes with two hydrogen bonds, not taking into account complexes with one hydrogen bond. In addition, the tautomeric forms of nucleic bases have not been considered in the present work.

\begin{figure}
	\begin{center}
		\resizebox{1\textwidth}{!}{%
			\includegraphics{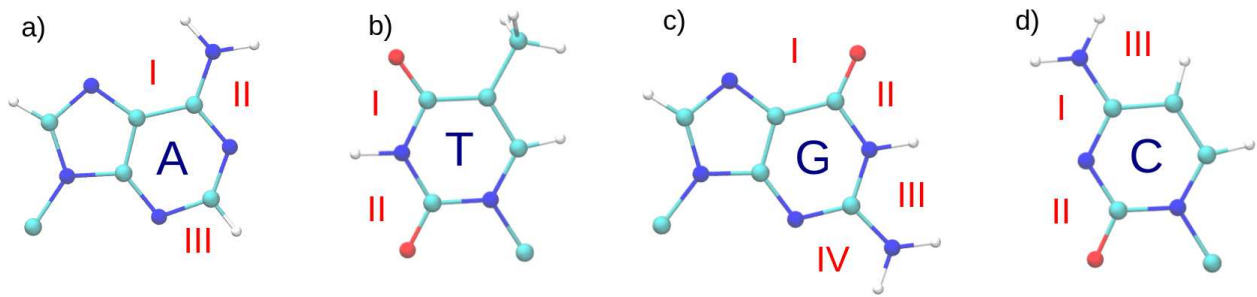}
		}
		\caption{The scheme of the sites of Adenine (a), Thymine (b), Guanine (c) and Cytosine (d) where the solvent molecule (water  or hydrogen peroxide molecule) can form a complex with a corresponding nucleic base with two hydrogen bonds. Roman numerals denote the number of the site.}
			\label{fig:sites} 
	\end{center}
\end{figure}

\begin{figure}
	\begin{center}
		\resizebox{0.8\textwidth}{!}{%
			\includegraphics{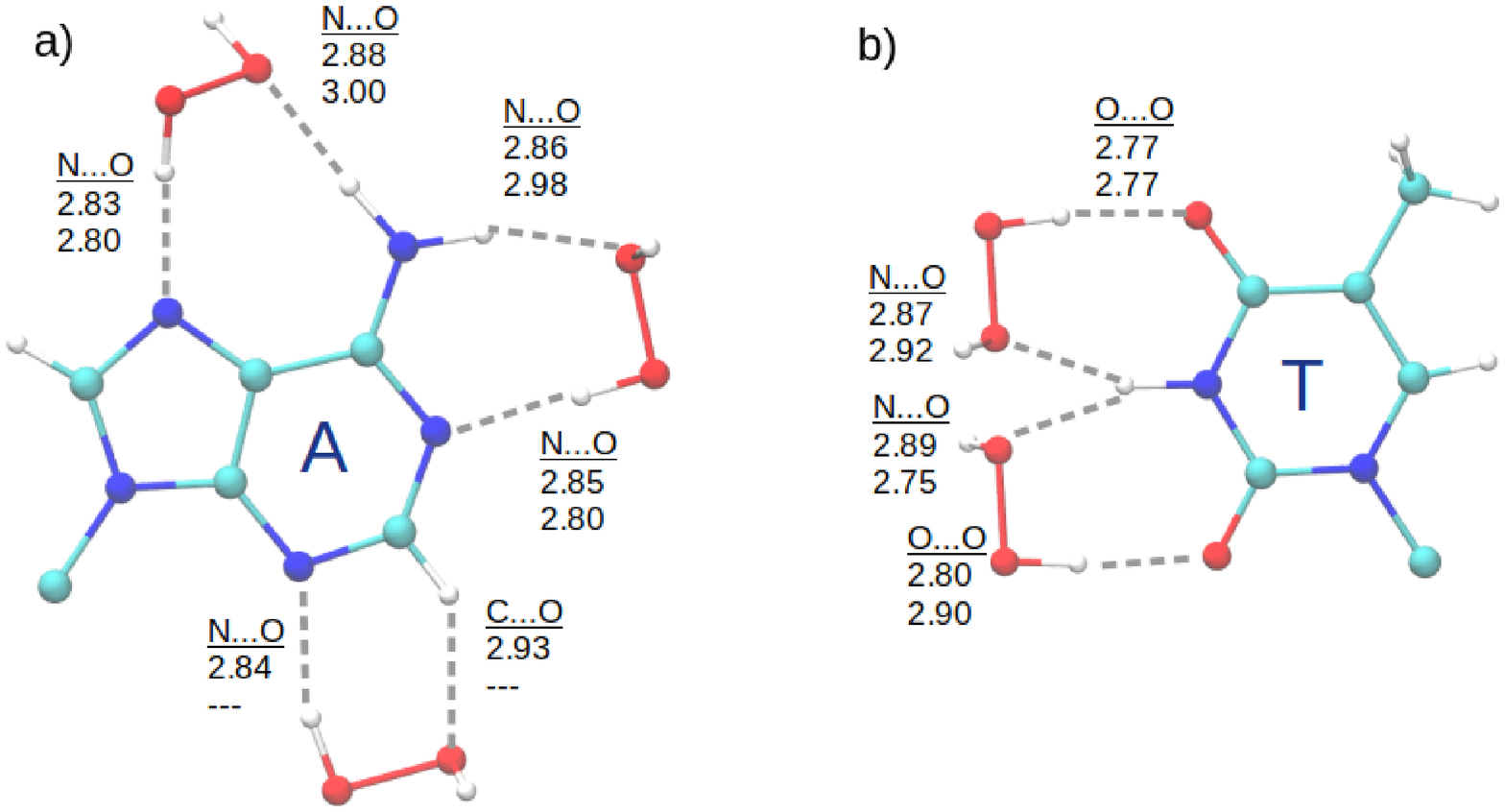}
		}
		\caption{Complexes of nucleic bases with hydrogen peroxide molecules: a) Adenine; b) Thymine. Numbers are given for the corresponding hydrogen bond distances between heavy atoms. Upper number corresponds to the value obtained in AAPFh method, and bottom number is obtained from B3LYP approach. }
		\label{fig:AT} 
	\end{center}
\end{figure}

\begin{figure}
	\begin{center}
		\resizebox{0.8\textwidth}{!}{%
			\includegraphics{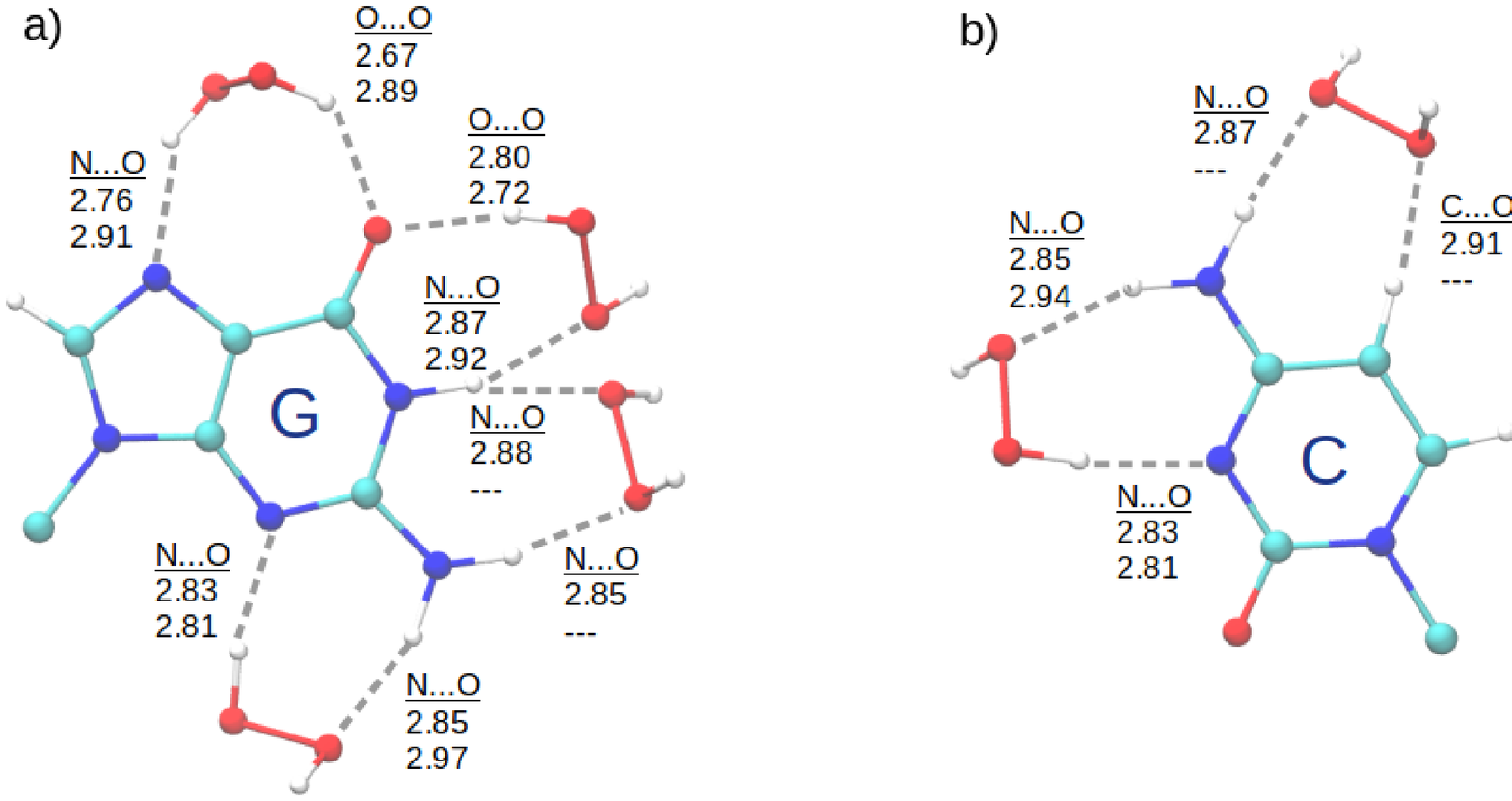}
		}
		\caption{Complexes of nucleic bases with hydrogen peroxide molecules: a) Guanine; b) Cytosine. Numbers are given for the corresponding hydrogen bond distances between heavy atoms. Upper number corresponds to the value obtained in AAPFh method, and bottom number is obtained from B3LYP approach. }
		\label{fig:GC} 
	\end{center}
\end{figure}

\begin{table}[th]
	\begin{center}	
		\noindent\caption{Interaction energies for Adenine, Thymine, Guanine and Cytosine complexes with solvent molecules ($H_2O$ or $H_2O_2$), as well as the difference between these energies ($\Delta E =|E_{H2O2}-E_{H2O}| $) calculated by AAPFh and B3LYP methods. Energies are given in $kcal/mol$. {\textquotedblleft}---{\textquotedblright} means that there is no minimum with two hydrogen bonds in this site.
			\label{tab:results}}
		{\begin{tabular}{@{}cccccccccc@{}} 			& & & & & & & & & \\
				& & \multicolumn{2}{c}{$H_2O_2$} & & \multicolumn{2}{c}{$H_2O$} & & \multicolumn{2}{c}{$\Delta E$} \\   \hline
				& & & & & & & & &	\\
				& & AAPFh & (B3LYP) & & AAPFh & (B3LYP) &  & AAPFh & (B3LYP) \\  \hline
				& & & & & & & & & \\
				& I & $-10.71$ &  ($-10.81$) & & $-7.95$ & ($-9.65$) & & $2.76$ & ($1.16$) \\
				Adenine	& II & $-9.22$ &  ($-11.19$) & & $-5.65$ & ($-9.00$) & & $3.57$ & ($2.19$) \\
				& III & $-6.76$ &  (---) & & --- & (---) & & --- & (---) \\ 
				& & & & & & & & &	\\ \hline
				& & & & & & & & &	\\
				Thymine	& I & $-9.26$ &  ($-10.39$) & & $-5.54$ & ($-8.64$) & & $3.72$ & ($1.75$) \\
				& II & $-8.63$ &  ($-11.17$) & & $-5.26$ & ($-9.29$) & & $3.37$ & ($1.88$) \\		
				& & & & & & & & &	\\ \hline
				& & & & & & & & &	\\
				& I & $-9.67$ &  ($-10.98$) & & $-9.06$ & ($-8.45$) & & $0.61$ & ($2.53$) \\
				Guanine & II & $-9.56$ &  ($-13.43$) &  & $-5.99$ & ($-11.62$) & & $3.57$ & ($1.81$) \\
				& III & $-10.36$ &  (---) & & $-6.35$ & (---) & & $4.01$ & (---) \\	
				& IV & $-9.02$ &  ($-9.83$) & & --- & ($-7.83$) & & --- & ($2.00$) \\		    
				& & & & & & & & &	\\ \hline
				& & & & & & & & &	\\
				& I & $-11.22$ &  ($-12.92$) & & $-6.82$ & ($-11.11$) & & $4.40$ & ($1.81$) \\
				Cytosine & II & --- &  ($-10.39$) & & $-7.40$ & (---) & & --- & (---) \\
				& III & $-7.62$ &  (---) & & --- & (---) & & --- & (---) \\				
				& & & & & & & & & 	\\ \hline
		\end{tabular}}
	\end{center}
\end{table}

For each nucleic base a few binding sites for the interaction with hydrogen peroxide  exist. These binding sites are schematically shown of Fig. \ref{fig:sites}. For each site the length of the hydrogen bonding and the energy minima are calculated. Binding sites from the backbone side are not considered.

One can see that for Adenine base at the sites A-I, A-II hydrogen peroxide molecule interacts with this nucleic base stronger than water molecule (see Tabl. \ref{tab:results}). For the cites A-I, A-II interaction energies are close to the values obtained in the work.\cite{dobado1999}  AAPFh method shows that in the cite A-III there is a weak binding of $H_2O_2$ molecule with bended hydrogen bonds, at the same time B3LYP method shows no binding at this cite. It should also be mentioned that in the case of B3LYP method water and hydrogen peroxide molecules change their shape insignificantly --- deformation energies do not exceed $0.5$  $kcal/mol$.

Our calculations show that in the case of Thymine there are two binding sites of T-I, T-II (Fig. \ref{fig:sites}b). The corresponding complexes with the hydrogen peroxide molecule are shown on Fig. \ref{fig:AT}b. For water molecules and hydrogen peroxide, binding takes place at the same sites of Thymine, but with a significant difference in the interaction energy (Tabl. \ref{tab:results}). Both our approaches (AAPF and B3LYP) give a dominance in interaction energy of complex with peroxide compared to the same complex with water molecule. In the case of B3LYP method the deformation energy of peroxide and water molecules plays insignificant role ($<0.3$  $kcal/mol$).

Complexes of binding of hydrogen peroxide molecules to Guanine (G-I, G-II, G-III, G-IV) are also shown on Fig. \ref{fig:sites}a. Differences in the interaction energies are given in Tabl. \ref{tab:results}. Studying these complexes by B3LYP method shows that in position I hydrogen peroxide molecule is significantly deformed ($2.5$  $kcal/mol$), what is the reason of the differences between hydrogen bond distances in AAPFh and B3LYP methods (Fig. \ref{fig:GC}a).

Cytosine has two binding sites (Fig. \ref{fig:GC}b). At the site C-I there is a binding of both of the water and hydrogen peroxide molecule, and besides hydrogen peroxide is more energetically favorable to bind (Tabl. \ref{tab:results}). At the place C-III there is a binding only of hydrogen peroxide. For the site C-II there no binding of hydrogen peroxide in the framework of AAPFh method due to the geometrical features of $H_2O_2$ molecule, which is rigid. As it can be seen from the structure of the complex, for B3LYP method in position II hydrogen peroxide dramatically changes its dihedral angle (deformation energy $4.2$  $kcal/mol$) to form two hydrogen bonds with Cytosine. This allows $H_2O_2$ molecule to make a stable complex with Cytosine.

Summarizing the results of our calculations by both methods used in the present work, it can be seen that for Adenine, Thymine, Guanine and Cytosine nucleic bases there are places which are more favorable for binding of hydrogen peroxide molecule compared to water molecule.

\section{Discussion and conclusions}
\label{discuss}
In the present work we have obtained the stable configurations of hydrogen peroxide  with DNA specific recognition sites by two methods: atom-atom potential functions and density functional theory. Both methods show similar results. It  can be seen (Fig. \ref{fig:AT} and \ref{fig:GC} ) that it is most probable for hydrogen peroxide to bind to Thymine from the side of complementary hydrogen bonds. To Adenine and Guanine - from the side of major and minor grooves as well as complementary hydrogen bonds. And to Cytosine - from the side of the major groove and the complementary hydrogen bonds.

As known,\cite{BookRecognition} the protein recognition of DNA macromolecule can take place as from the major as well as from the minor grooves, depending on the type of enzyme and on the form of double helix. Therefore,
the formation of complex of a nucleic acid with hydrogen peroxide molecule, which binds to the base from the major
(Fig. \ref{fig:sites}, A-I, C-III, G-I) or minor groove (Fig. \ref{fig:sites}, A-III, G-IV), can prevent the recognition of this base by the enzyme and, therefore, block the process of protein-nucleic recognition.

The formation of the same complexes from the side of complementary hydrogen bonds (TI, T-II, A-II, CI, C-II, G-II, G-III) can take place at the stage of the DNA transcription process, when the double-stranded DNA
is already unzipped up to two single strands. In this case, the energy of nucleic base blocking by hydrogen peroxide molecule is comparable to the energy of the complementary pairs formation.\cite{ourWorkUnz}

To sum up, depending on the form of DNA, binding to the DNA bases in the solution can occur both from the sides of major and minor grooves. By this means there are sites of nucleic bases, where binding to hydrogen peroxide is much more advantageous than that of a water molecule. It should be mentioned, that when considering interaction with active sites of DNA backbone,\cite{pyatEPJ2015} the sufficient preference of hydrogen peroxide compared to water molecule was not found. In the present work there are sites of nucleic bases which are much more favorable for $H_2O_2$ molecule to bind.  Consequently, the formation of such complexes is sufficiently probable in cell during the ion beam therapy treatment, and in this way  the processes of genetic information transfer should be blocked in cancer cells.

\section{Acknowledgments}
The present work was partially supported by by the program “Grid infrastructure and Grid technologies for scientific and applied applications”  of the National Academy of Sciences of Ukraine (project number 0118U000662). 

The materials of this work were presented at Belgrade Bioinformatics Conference 2018. One of the authors (S.N.V.) is grateful to the organization committee of BELBI 2018 for the financial support and hospitality.

\end{document}